\documentclass[aps, prl, 10pt, twocolumn, superscriptaddress, noshowpacs, preprintnumbers]{revtex4-2}

\usepackage[colorlinks=true,breaklinks=true]{hyperref}
\hypersetup{allcolors=[rgb]{0.05 0.05 0.75},linkcolor=[rgb]{0.75 0.05 0.05}}
\usepackage[dvipsnames]{xcolor}
\usepackage{mathtools}
\usepackage{amsmath}
\usepackage{amsfonts}
\usepackage{amssymb}
\usepackage{bbold}
\usepackage{bm}
\usepackage{orcidlink}
\usepackage{esint}
\usepackage{microtype}
\usepackage{booktabs}
\usepackage{float}

\long\def\exclude#1{}

\newcommand{\edit}[1]{{\color{black}{#1}}}

\newcommand{\bK}{{\bf K}}
\newcommand{\bk}{{\bf k}}
\newcommand{\bp}{{\bf p}}
\newcommand{\br}{{\bf r}}
\newcommand{\bu}{{\bf u}}
\newcommand{\bv}{{\bf v}}
\newcommand{\bD}{{\bf D}}
\newcommand{\bkappa}{{\boldsymbol{\kappa}}}
\newcommand{\GF}{G_{\rm F}}
\newcommand{\sH}{{\sf H}}
\renewcommand{\sp}{{\sf p}}

\begin{document}

\title{Flavomons in Matter Gradients: Ray Tracing and Amplitude Evolution
}

\author{Damiano F.\ G.\ Fiorillo \orcidlink{0000-0003-4927-9850}}
\affiliation{Istituto Nazionale di Fisica Nucleare (INFN), Sezione di Napoli,
Complesso Universitario di Monte Sant’Angelo, Via Cintia, 80126 Napoli, Italy}
\affiliation{Gran Sasso Science Institute (GSSI), Viale Francesco Crispi 7, 67100 L’Aquila,
Italy}

\author{Georg G.\ Raffelt
\orcidlink{0000-0002-0199-9560}}
\affiliation{Max-Planck-Institut f\"ur Physik, Boltzmannstr.~8, 85748 Garching, Germany}

\begin{abstract}
Flavor instabilities develop in neutrino plasmas through the emission of flavomons, collective excitations of the flavor field. 
Matter gradients can sweep flavomons across the unstable wave-number range, halting their amplification. This effect is parametrically small for fast modes, where it was first identified, but we show that it strongly alters neutrino-mass-induced instabilities below the supernova shock, while leaving those outside largely unaffected. Dispersion relations alone, which are inevitably local, cannot determine the cumulative amplification of slow instabilities.
Using the WKB approximation, we derive the covariant evolution equation for the flavomon amplitude along geometrical-optics rays, enabling a global treatment of instability growth and seeding.
\end{abstract}

\maketitle

{\bf\textit{Introduction}}---The evolution of core-collapse supernovae (SNe) hinges on neutrinos, which dominate energy and lepton-number transport and are central to the explosion mechanism, nucleosynthesis, and observational tests \cite{Bethe:1985sox, Mirizzi:2015eza, Burrows:2020qrp, Boccioli:2024abp, Janka:2025tvf, Jerkstrand:2025, Raffelt:2025wty}. A key challenge for theory and simulations is neutrino--neutrino refraction \cite{Pantaleone:1992eq}, which causes collective flavor evolution \cite{Duan:2009cd, Duan:2010bg, Tamborra:2020cul, Richers:2022zug, Johns:2025mlm} through flavor instabilities. 

These instabilities are naturally understood as neutrinos emitting flavor waves \cite{Fiorillo:2024bzm,Fiorillo:2024pns}, or equivalently as the stimulated emission of their quanta, flavomons \cite{Fiorillo:2025npi}. The kinematics of this process, together with flavor lepton-number conservation, determines the instability conditions in both the ``fast'' regime, the limit of massless neutrinos \cite{Fiorillo:2024bzm,Fiorillo:2024uki,Fiorillo:2025npi}, and the ``slow'' regime, where neutrino masses are essential \cite{Fiorillo:2024pns,Fiorillo:2025ank,Fiorillo:2025zio,Fiorillo:2025kko}. \edit{On a formal level, Johns and Kost~\cite{Johns:2025yxa}  have also advocated treating flavor waves as effective degrees of freedom, a key ingredient for SN flavor evolution.}

The properties of flavor waves follow from their dispersion relation \cite{Izaguirre:2016gsx,Capozzi:2017gqd,Martin:2019gxb}, assuming a uniform environment. The wavelengths of unstable modes, set by the neutrino--neutrino interaction strength $\mu=\sqrt{2}\GF n_\nu$, are of order centimeters. Despite this small scale, inhomogeneities can impede instability growth \cite{Sigl:2021tmj, Bhattacharyya:2025gds, Fiorillo:2025gkw}. The matter potential $\lambda=\sqrt{2}\GF n_e$ far exceeds $\mu$ within the SN shock-wave radius, so its spatial variation can become important.


This effect was \edit{numerically anticipated in a schematic study} \cite{Sigl:2021tmj} and \edit{later more clearly} identified in a two-beam model of fast flavor evolution \cite{Bhattacharyya:2025gds}. An exact solution showed that strong matter gradients can prevent instability growth altogether, \edit{although, under realistic conditions, the impact is parametrically small.} \edit{However, Ref.~\cite{Bhattacharyya:2025gds} speculated that it can be more dramatic for slow instabilities}, whose growth rates are small despite their short wavelengths. As the earliest to appear in SNe~\cite{Fiorillo:2025gkw}, slow instabilities could modify the entire subsequent evolution, making their possible suppression by matter gradients particularly consequential.

In this Letter, we develop a practical framework for flavor instabilities in nonuniform media. Matter gradients advect flavomons through wave-number space, so that unstable modes can leave the instability band before significant growth. Gravitational redshift has a similar but subdominant effect near a SN core.

\edit{The matter-gradient effect of Refs.~\cite{Sigl:2021tmj, Bhattacharyya:2025gds} becomes transparent through geometrical optics. The role of group velocity for flavomon propagation was recognized in Ref.~\cite{Fiorillo:2025ank}, while Ref.~\cite{Johns:2025yxa} later advocated more generally for a geometrical-optics approach, \edit{postulating} an amplitude-transport equation. Here, we derive the geometrical-optics transport from the underlying equations of motion. Our result has the standard WKB structure
\cite{thorne2017modern}, but differs from the \edit{incomplete} prescription of Ref.~\cite{Johns:2025yxa}.}

Most importantly, we quantify how matter gradients counteract slow instabilities. Within tens of milliseconds of SN bounce, such gradients strongly inhibit growth below the shock wave, although complete suppression is uncertain. Beyond the shock, gradients are too weak to prevent growth, allowing slow instabilities to shape the flavor composition of the emerging neutrino flux.

The gradient-imposed limit to instability growth makes physical seeding essential. Quantum fluctuations provide an irreducible source through spontaneous flavomon emission, but neutrino mixing is expected to dominate in practice. Our covariant amplitude-transport equation incorporates the mixing-induced source, enabling a global description of linear instability growth.


{\bf\textit{Dispersion relation}}---Neutrino flavor information is encoded in the $3\times3$ density matrix $\varrho_\bp(\br,t)$ associated with each momentum $\bp$. We restrict our discussion to two flavors, although the framework of flavor waves can be extended to three flavors~\cite{Fiorillo:2025npi}. For two flavors, we write
\begin{equation}\label{eq:rho-expansion}
    \varrho_\bp=\frac{1}{2}\begin{pmatrix}
        n_\bp+D_\bp&\psi^*_\bp\\
        \psi_\bp & n_\bp-D_\bp
    \end{pmatrix},
\end{equation}
where $n_\bp=f_\bp^{\nu_e}+f_\bp^{\nu_\mu}$ is the total occupation number,  $D_\bp=f_\bp^{\nu_e}-f_\bp^{\nu_\mu}$ is the DLN, or electron--muon lepton number difference,  and $\psi_\bp(\br,t)$ the field of flavor coherence. In the homogeneous and collisionless limit, $\varrho_\bp$ follows the usual kinetic equation \cite{Dolgov:1980cq, Rudsky, Sigl:1993ctk, Fiorillo:2024fnl, Fiorillo:2024wej} 
\begin{equation}\label{eq:QKE}
    (\partial_t+\bv\cdot\partial_\br)\varrho_\bp=-i\left[\sH_\bp,\varrho_\bp\right],
\end{equation}
where for ultrarelativistic neutrinos, $\bv=\bp/|\bp|$, and the Hamiltonian is $\sH_\bp=\sH_{\bp}^{\rm vac}+\sH_{\bp}^{\rm mat}+\sH_{\bp}^{\nu}$.

To include antineutrinos, we use the flavor-isospin convention, where $\nu_e\to\overline{\nu}_\mu$ and $\nu_\mu\to \overline{\nu}_e$, and $\overline\nu$ modes are represented as  $\nu$ modes with negative $E$. Overall phase space is covered by ${\sf p}=\{E,\bv\}$ with $\bv$ on the unit sphere and $-\infty<E<+\infty$. The vacuum oscillation frequency $\omega_E=\delta m^2/2E$ (mass splitting $\delta m^2$) is positive or negative depending on the sign of $E$ and on the mass ordering, which is encoded in the sign of $\delta m^2$.

In the weak-interaction basis, the vacuum oscillation term in Eq.~\eqref{eq:QKE} is \smash{$\sH_\sp^{\rm vac}=\frac{1}{2}\omega_E( s_V \sigma_1 -c_V\sigma_3)$}, with $\sigma_i$ Pauli matrices, $c_V=\cos 2\theta_V$,  $s_V=\sin 2\theta_V$, and $\theta_V$ the vacuum mixing angle. Matter refraction contributes \smash{$\sH_\sp^{\rm mat}=\frac{1}{2} \lambda(1-\bu\cdot\bv)\sigma_3$}, where $\lambda=\sqrt{2} \GF n_e$ and $\bu$ is the matter bulk velocity. Neutrino--neutrino refraction contributes
\begin{equation}
        {\sf{H}}_\sp^{\nu}=\mu\sum_{\sp'} \varrho_{\sp'}(1-\bv\cdot\bv').
\end{equation}
Here $\sum_{\sp}=n_\nu^{-1}\int_{-\infty}^{+\infty}E^2 dE\int d^2\bv/(2\pi)^3$ means a phase-space integral with $\bv$ on the unit sphere, where $n_\nu$ is the total neutrino density and $\mu=\sqrt{2}G_F n_\nu$ is the typical neutrino--neutrino interaction strength.

Linearizing Eq.~\eqref{eq:QKE} in $\psi_\sp(\br,t)$ yields the dispersion relation for flavor waves \cite{Izaguirre:2016gsx, Capozzi:2017gqd, Martin:2019gxb}. Without repeating details, we follow our earlier papers \cite{Fiorillo:2024pns, Fiorillo:2025zio} and introduce the neutrino four-velocity $v^\mu=(1,\bv)$, the flavomon four-momentum $K^\mu=(\Omega_\bK,\bK)$, and its shifted version $k^\mu=K^\mu+D^\mu+\lambda^\mu=(\omega_\bk,\bk)$ in terms of the angular moments $D^\mu=\sum_\sp D_\sp v^\mu$ and $\lambda^\mu=\lambda(1,\bu)$. With these definitions, the unstable eigenmodes are 
\begin{equation}
    \psi_\sp=\frac{\psi\cdot v}{k\cdot v-\omega_E c_V}\mu D_\sp,
\end{equation}
where $\psi^\mu=\sum_\sp \psi_\sp v^\mu$, and $\psi\cdot v=\psi_\mu v^\mu$ and so forth.

Self-consistency of the EoMs in Fourier space yields the dispersion relation in the form
$\varepsilon_{\mu\nu}\psi^\nu=0$ with the dielectric tensor
\begin{equation}\label{eq:dielectric_tensor}
    \varepsilon_{\mu\nu}=g_{\mu\nu}-\sum_\sp \frac{\mu D_\sp v_\mu v_\nu}{k\cdot v-\omega_E c_V}.
\end{equation}
Notice that  $s_V$, encoding the off-diagonal mass term, does not appear and acts only as a seed for the instability. 

We will not need a more explicit form of the dispersion relation. The frequency $\omega_\bk$ depends only on~$\bk$, with matter effects absorbed into the shifted four-momentum. For the physical flavomon frequency, one finds
\begin{equation}\label{eq:frequency}
    \Omega_\bK=\omega_{\bk}-\lambda-\mu D_0
\end{equation}
with 
$\bk=\bK+\mu\bD_1+\lambda \bu$.

{\bf\textit{Flavomon equations of motion}}---Inhomogeneities imply that neutrinos experience external forces. Near a SN core, gravity strongly dominates over the refractive force caused by matter gradients. Formally, the kinetic equation~\eqref{eq:QKE} acquires an additional momentum-drift term on the left-hand side \cite{Sigl:1993ctk}. Throughout this work, we assume that the prescribed local neutrino distribution already incorporates these effects.

Flavomon trajectories, in contrast, are governed by the local dispersion relation of the slowly varying medium. The geometric-optics limit applies because the cm-scale flavomon wavelengths, while macroscopic, remain much smaller than the multi-km scale of background variations. Gravitational effects are parametrically small because they scale with the flavomon energy, which is itself of the same order as the matter potential $\lambda$. A formal derivation is given in the Supplemental Material (SM) \cite{Supplemental-Material}.

Assuming a stationary medium with local flavomon dispersion relation $\Omega(\bK,\br)$, the standard geometric-optics equations are
\begin{equation}
    \frac{d\br}{dt}=\frac{\partial \Omega_\bK}{\partial\bK}
    \quad\text{and}\quad  
    \frac{d\bK}{dt}=-\frac{\partial \Omega_\bK}{\partial \br}.
\end{equation}
These are textbook results for wave propagation in non\-uniform media, also stated specifically for flavor waves in Ref.~\cite{Johns:2025yxa}.
As the matter potential $\lambda$ dominates the dispersion relation, one finally finds
\begin{equation}\label{eq:eq_motion_flavomons}
    \frac{d\br}{dt}=\bv_{\rm gr}
    \quad\text{and}\quad 
    \frac{d\bK}{dt}=\frac{\partial \lambda}{\partial \br}(1-\bv_{\rm gr}\cdot\bu),
\end{equation}
where $\bv_{\rm gr}=\partial_\bK \Omega_\bK$ is the group velocity. In the absence of a matter flow, the term proportional to $\bu$ disappears. \edit{For a weakly unstable mode, these equations refer only to the real part of the eigenfrequency determining the evolution of the wave phase; the growth rate only enters the amplitude evolution.}

The matter gradient usually deflects the ray, but also changes \(|\mathbf K|\), even for propagation along the gradient. If the instability is confined to a narrow range of $|\mathbf K|$, the wave packet quickly leaves this band, limiting its growth. This effect was observed in the two-beam fast-flavor model of Ref.~\cite{Bhattacharyya:2025gds}, although for an unrealistically large matter gradient. It is parametrically small for fast modes but can severely limit the growth of slow instabilities.


{\bf\textit{Fast instabilities}}---Momentum drift limits growth only for very weak instabilities. Considering an axially symmetric system, a weak fast instability is driven by a shallow angular crossing. The DLN is positive over all directions, except for a narrow interval, $\Delta v=\Delta \cos\theta$, where a small negative DLN exists (``flipped neutrinos''), whose typical size is measured by $\Delta D_v$. \edit{Notice that, while neutrino--neutrino interactions are measured by $\mu$, which depends on the total number density, it is only the DLN $D_\sp$ that enters flavor evolution: therefore, if we denote its typical order of magnitude as $\epsilon$, the typical scale of fast instability depends on $\mu \epsilon$.}


The instability mechanism, elucidated in Refs.~\cite{Fiorillo:2024bzm, Fiorillo:2024uki}, consists of flavomon emission by neutrinos satisfying the massless Cherenkov condition $k\cdot v=0$. The growth rate scales with the flipped-neutrino population, $\gamma\sim \mu\Delta D_v$. The resonance condition implies an unstable wave-number range $\Delta K\sim \mu \epsilon \Delta v$, as also seen from Eqs.~(5.27) and~(5.32) of Ref.~\cite{Fiorillo:2024uki}. These estimates apply to sufficiently weak instabilities
that the frequency width $\gamma$ does not broaden the resonance across the entire unstable range. Thus, $\gamma/\mu\epsilon\ll \Delta v$, or $\Delta D_v\ll  \epsilon \Delta v$.

For axisymmetric distributions, three unstable \(K\) ranges appear, corresponding to two longitudinal modes and one transverse mode \cite{Yi:2019hrp, Fiorillo:2024uki, Fiorillo:2024dik}. Their comparable widths leave the argument unchanged.

Under such circumstances, the growth rate and the unstable $K$ range may be quite small. If a wavepacket begins within this range, $K$ evolves by the matter gradient. According to Eq.~\eqref{eq:eq_motion_flavomons}, after a time $\delta t$, the change is $\delta K\sim \delta t\partial\lambda/\partial r$. Therefore, after $\Delta t=\Delta K/\partial_r \lambda$, the flavomon has left the unstable $K$ range and stops growing, limiting its final amplitude. The maximum number of e-folds is $N_{\text{e-folds}}\sim \gamma \Delta t\sim \mu^2 \epsilon \Delta D_v\Delta v/\partial_r\lambda$. If it does not even reach~1, there is no meaningful growth. Therefore, the parametric condition to neglect instability quenching by matter gradients is $N_{\text{e-folds}}\gg1$ or
\begin{equation}\label{eq:matter_gradient_criterion_fast}
    \partial_r\lambda \ll \mu^2\epsilon \Delta D_v \Delta v.
\end{equation}
Whether this condition is met in practice depends on the strength of the crossing. Below the SN shock wave, a typical refractive potential is $\lambda \sim 10$--$100 \mu \epsilon$ that changes over a scale $\ell \sim 10\, \mathrm{km}\sim 10^6 (\mu\epsilon)^{-1}$ for typical $\mu\epsilon\sim 1\,\mathrm{cm}^{-1}$. Therefore, matter gradients can be neglected for instabilities so strong that $\Delta v\,\Delta D_v/D_v\gg 10^{-4}$.

While our finding agrees with Ref.~\cite{Bhattacharyya:2025gds}, which also found a time-dependent $K$ for a uniform matter gradient, we explain this effect through our general geometrical-optics approach. Moreover, Ref.~\cite{Bhattacharyya:2025gds} uses a two-beam model, for which instabilities are not resonant. Our parametric expressions hold for a continuous angular distribution, which is needed to examine very weak instabilities \cite{Fiorillo:2024uki}, as these are the ones sensitive to matter gradients.

{\bf\textit{Slow instabilities}}---To study how matter gradients limit their growth, we consider a SN core a few tens of milliseconds post bounce, below the shock wave, but far beyond the proto-neutron star, where slow instabilities first appear~\cite{Fiorillo:2025gkw}. We model the distributions as pinched quasi-thermal spectra---for details see the~SM~\cite{Supplemental-Material}. 

\begin{figure}
    \centering
    \includegraphics[width=\columnwidth]{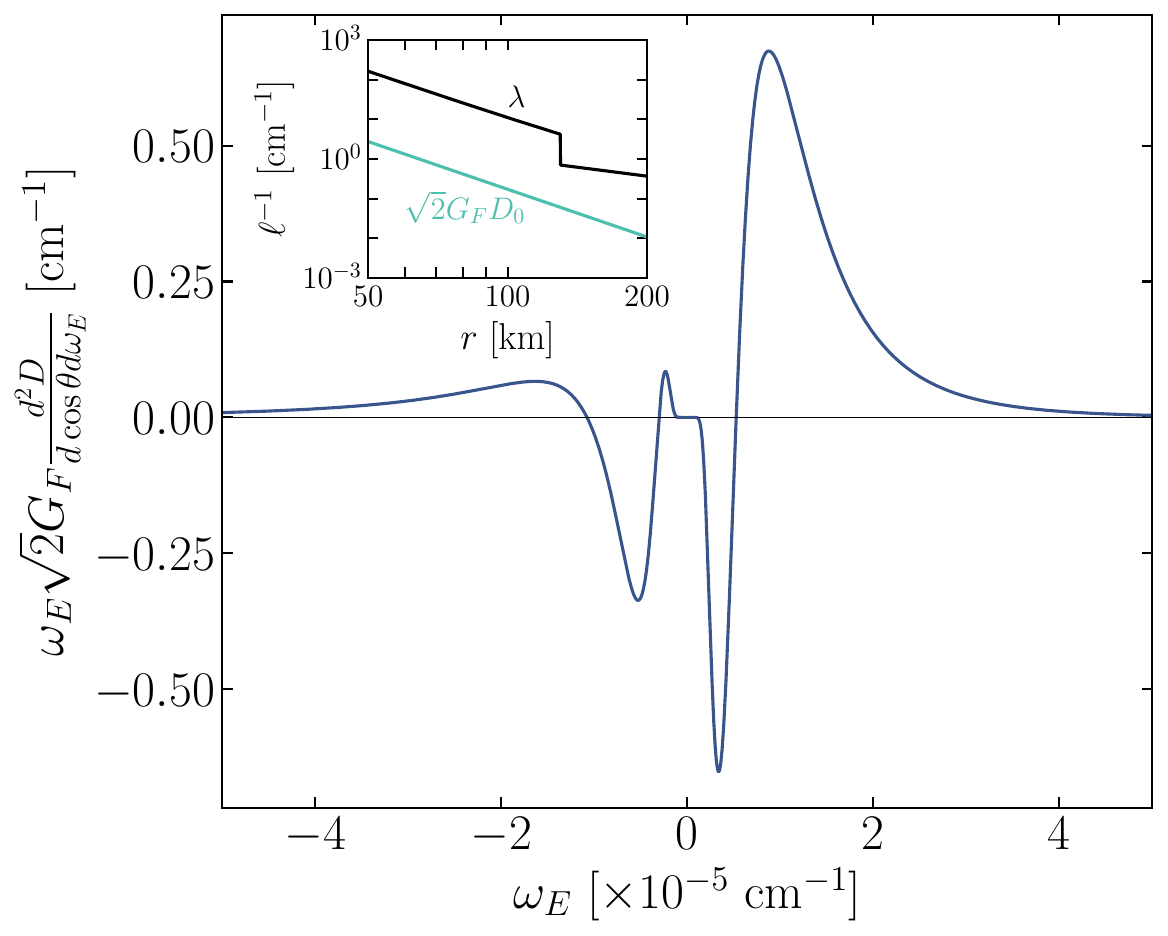}
    \caption{Schematic DLN distribution for neutrinos moving in the radial direction in the region below the SN shock wave, tens of milliseconds after core bounce. The inset shows schematic profiles for the matter and neutrino refractive potentials, the latter measured by the total DLN density $D_0=\sum_\sp D_\sp$.}
    \label{fig:neutrinos}
\end{figure}

Figure~\ref{fig:neutrinos} shows a typical DLN distribution for the radial direction as a function of $\omega_E=\delta m^2 V/2E$ (we set for simplicity $c_V=1$), where negative $\omega_E$ pertain to antineutrinos if $\delta m^2>0$. For large positive $\omega_E$, low-energy $\nu_e$ dominate, whereas for large negative $\omega_E$, it is $\overline\nu_\mu$ that dominate, both providing positive DLN because $\nu_e$ and $\overline\nu_\mu$ have the same flavor isospin. For small $\omega_E$, large-energy $\nu_\mu$ overshoot $\nu_e$, providing a negative DLN dip, and analogous for small negative $\omega_E$, where $\overline\nu_e$ overshoot. This range provides most of the flipped population that emits flavomons and thus causes the mass-driven instability.

We have previously examined the main properties of neutrino-mass-driven instabilities in detail \cite{Fiorillo:2024pns, Fiorillo:2025ank, Fiorillo:2025zio, Fiorillo:2025gkw, Fiorillo:2025kko}. Typical growth rates reach $\gamma\sim \omega_E/\epsilon$, again with $\epsilon$ the typical asymmetry in the DLN between $\nu$ and $\overline\nu$. Our schematic conditions roughly yield $\omega_E\sim 10^{-5}\,\mathrm{cm}^{-1}$ and $\epsilon\sim 0.1$, so we expect $\gamma\sim 10^{-4}\,\mathrm{cm}^{-1}$. A numerical solution of the dispersion relation (see SM~\cite{Supplemental-Material}) fully confirms this conclusion. Unstable modes extend to a $K$ range of order $\Delta K\sim \mu \epsilon\sim 0.1\,\mathrm{cm}^{-1}$, also confirmed by our numerical solution in the SM~\cite{Supplemental-Material}.

We now study the impact of a matter gradient. In the inset of Fig.~\ref{fig:neutrinos}, we show the schematic behavior of the matter and neutrino refractive potentials. Below the shock wave, matter strongly dominates, as expected, with a typical $\lambda\sim 10\,\mathrm{cm}^{-1}$, varying over a scale $\ell \sim 100\,\mathrm{km}$. We therefore examine the evolution of an unstable flavomon under the EoMs~\eqref{eq:eq_motion_flavomons}. We assume $\lambda \gg \mu \epsilon$ everywhere.

For short timescales, it is mainly the flavomon momentum that evolves, $d\bK/dt=\partial_\br \lambda=(\partial_r\lambda)\br/r$ for a purely radial matter gradient, whereas spatial propagation is neglected because the group velocity of slowly-unstable flavomons is comparable with the speed of light~\cite{Fiorillo:2024pns, Fiorillo:2025ank, Fiorillo:2025zio}, so the flavomons move little during a growth timescale. However,
the wavevector changes considerably as
\begin{equation}\label{eq:wavevector_local}
    \bK=\bK(t=0)-\left|\frac{\partial \lambda}{\partial r}\right|\frac{\br}{r}\,t.
\end{equation}
While this is the same mechanism as for fast modes, the growth now can be limited far more efficiently. Since the instability extends only to $\Delta K\sim \mu \epsilon$, it takes only $\Delta t\sim \Delta K/|\partial_r\lambda|$ to transit through the unstable $K$ range. Based on our schematic model, a maximum number $N_{\text{e-folds}}\sim \gamma \Delta t\sim \mu \omega_E \ell/\lambda\sim 1$--10 can be accumulated. Therefore, below the shock wave, the matter gradient dramatically limits the maximum development of the instability.

The fate of the unstable wavepackets is hard to assess without a more detailed analysis in a realistic situation. The wavevector is rapidly reduced by the matter gradient, but we cannot unambiguously conclude that it will turn negative. First of all, in leaving the unstable branch, the mode may pass through a near-crossing point of two near-stable branches (see, e.g., the discussion in Ref.~\cite{Fiorillo:2026zcm}). Such points are found for weakly anisotropic distributions, so it remains to be seen how general they are. 

At crossing points, the validity of the WKB approximation requires careful assessment because nearby branches may invalidate the adiabatic approximation.
Second, if the distribution is strongly anisotropic, the backward-moving neutrino population may simply be too small, so that backward-moving flavomons may not be supported. The global evolution of slow modes definitely must be explored, but depends on first advancing our understanding of their dispersion relation.

Beyond the shock wave, the situation is qualitatively different. We assume a density jump by roughly a factor of~5, consistent with expectations from Rankine-Hugoniot conditions~\cite{landau1987fluid}. Outside of the shock wave, the density drops roughly as in hydrodynamic free fall, with \smash{$\lambda\propto r^{-3/2}$}. 
The gradient is then smaller by more than an order of magnitude.
While $\mu\epsilon\ll \lambda$ still holds, 
the timescale for wavevector drift increases by an order of magnitude, 
and the number of e-folds grows in proportion. This effect is even stronger at larger distances, where the gradient decreases in proportion to the radius. 
Thus, matter gradients still affect slow instabilities outside the shock wave but do not prevent their growth.

\edit{{\bf\textit{Amplitude evolution and seeding}}---The number of e-folds accrued below the shock wave is too small to guarantee saturation, but large enough to make it conceivable. Local stability analysis alone cannot settle this question: we must follow the amplitude evolution.

In geometrical optics, we express the zero-component of the flavor field as $\psi^0=a e^{i S}$, where $a(t,\br)$ is a slowly varying amplitude and $S(t,\br)$ a rapidly varying eikonal. Conservation of flavomon number (or wave action) along the ray suggests using instead $\Psi=\psi^0/\sqrt{\mathcal{Z}}$ through the wavefunction renormalization $\mathcal{Z}^{-1}=n_\mu \partial_\Omega \varepsilon^{\mu\nu} n_\nu$. Here,  $n_\mu$ is the eigenvector of our specific mode, normalized such that $n^0=1$. The factor $\mathcal{Z}$ depends on the local dispersion relation and wavevector $\bK(t,\br)=\boldsymbol{\nabla} S(t,\br)$. Neglecting growth and seeding, flavomon-number conservation gives $\partial_t |\Psi|^2+\boldsymbol{\nabla}\cdot (|\Psi|^2 \bv_{\rm gr})=0$. A purely classical but lengthier derivation is provided in the SM~\cite{Supplemental-Material}.

From each point in space at $t=0$, we follow the trajectory of a wave with initial wavevector $\bK_0(\br)=\boldsymbol{\nabla}S(0,\br)$, defining a ray $\{\br(t),\bK(t)\}$. Along each such ray, the phase advances by $\delta S=(\bv_{\rm gr}\cdot \bK-\Omega_R)\delta t$ over a short interval
$\delta t$. The normalized amplitude $\mathcal{A}=a/\sqrt{\mathcal{Z}}$ evolves through the continuity equation and grows as $e^{\Omega_I t}$, so that
\begin{equation}\label{eq:transport_flavomons}
    \frac{d\Psi}{dt}+\frac{1}{2}\boldsymbol{\nabla}\cdot \bv_{\rm gr}\Psi=i(\bK\cdot \bv_{\rm gr}-\Omega)\Psi+\mathcal{S}_{\bK}.
\end{equation}
For a given ray, the amplitude evolves as a function of time, with the total derivative $d/dt=\partial_t+\bv_{\rm gr}\cdot \boldsymbol{\nabla}$. The divergence term describes ray focusing or defocusing, thus modifying the amplitude. Analogous results are standard in geometrical optics, see e.g.~Sec.~7.3 of Ref.~\cite{thorne2017modern}. 

The last term describes instability seeding by neutrino phase-space density fluctuations $\delta D_{\mathbf p}$. We take the local Fourier component $\delta D_{\mathbf p,\mathbf K}$ of these fluctuations in a region around each ray point, large compared with $|\mathbf K|^{-1}$ but small compared with the background variation scales. Projecting the flavor-mixing source gives
\begin{equation}\label{eq:seeding}
    \mathcal{S}_\bK=-i\sqrt{\mathcal{Z}}\sum_{\sp}\frac{\mu \omega_E s_V \delta D_{\sp,\bK}v\cdot n}{k\cdot v-\omega_E c_V}.
\end{equation}
A derivation is provided in the End Matter.

Equation~\eqref{eq:transport_flavomons} evolves the normalized field along a specified ray, rather than a distribution over independent wavevectors.  Its amplitude evolution follows standard WKB theory, including wave-action conservation and the resulting focusing or defocusing of rays. This differs from the amplitude-transport prescription in Eqs.~(88)--(90) of Ref.~\cite{Johns:2025yxa}, which do not incorporate these effects. Our equation also includes the explicit seeding term $\mathcal S_{\mathbf K}$.

Once the amplitude has grown enough that further seeding is negligible, we can follow the occupation-number distribution $N(\br,\bK;t)=|\Psi|^2\delta[\bK-\tilde{\bK}(\br,t)]$, where $\tilde{\bK}$ is the local wavevector of the ray family, obeying
\begin{equation}\label{eq:N-equation}
\partial_t N
+\frac{d\br}{dt}\cdot\partial_\br N
+\frac{d\bK}{dt}\cdot\partial_\bK N
=2\gamma_{\bK}N,
\end{equation}
where $\gamma_{\bK}=\Omega_I$. Here, $d\mathbf r/dt$ is the group velocity, while $d\mathbf K/dt$ describes wavevector drift. Coupled to the neutrino kinetic equation in quasi-linear theory \cite{Fiorillo:2024qbl, Fiorillo:2025npi, Fiorillo:2026byh}, this equation describes an interacting inhomogeneous plasma of neutrinos and flavomons. The amplitude equation is needed to include seeding, while subsequent evolution can be followed through occupation numbers.
Geometrical focusing is now absent since focusing in 
coordinate space is compensated by defocusing in wave-numbers. 
}

{\bf\textit{Discussion}}---Applying geometric-optics methods to flavor waves, we have developed a ray-tracing framework for instability growth. Our key finding is that wavevector drift caused by matter gradients can strongly limit their amplification. This effect explains physically the earlier result for a simplified fast flavor system \cite{Bhattacharyya:2025gds}, while revealing that parametrically it is a small effect for fast modes. For slow instabilities, however, one must go beyond local stability analysis and follow their evolution along rays.

We have accounted only for radial matter gradients. Nonradial gradients generated by turbulence would make the impact even larger, but we refrain from far-reaching speculations. The properties of sub-grid turbulent density fluctuations in SNe are not widely explored and may not satisfy the usual conditions for Kolmogorov turbulence.

Our results have far-reaching consequences for flavor evolution in SNe. Outside the shock radius, WKB drifts are unlikely to prevent saturation of slow instabilities even at early post-bounce times. As the first instabilities to appear, they may well shape early flavor evolution. Below the shock radius, slow modes are drastically affected, yet growth is not inhibited enough to ignore them. 

Therefore, one needs to go beyond conventional linear stability analysis and follow amplitude evolution. We provide a new ray-tracing method, not evolving Fourier amplitudes in time, but rather following flavomon trajectories. While the use of geometrical optics was anticipated in Refs.~\cite{Fiorillo:2025ank,Johns:2025yxa,Fiorillo:2025gkw}, here we cast it into a practical framework to follow instability growth and~seeding.

The validity of WKB in the nonlinear phase needs to be scrutinized, since an unstable mode may branch into multiple stable ones---see the examples of Refs.~\cite{Fiorillo:2024uki, Fiorillo:2024dik} for the fast case, and~\cite{Fiorillo:2024pns, Fiorillo:2025ank, Fiorillo:2025zio} for the slow case. Similar concerns on branching points arise even in homogeneous setups with slowly evolving flavor waves~\cite{Liu:2025muc}.

A subtle point is that gravitational drifts for flavomons are subdominant, but not by much, as we show in our SM~\cite{Supplemental-Material}. This is not obvious because for neutrinos, gravitational drifts dominate by many orders of magnitude. The difference is that the total flavomon energy is of order the weak-interaction potential, whereas neutrino energies are many orders of magnitude larger.

With much of the focus on fast flavor instabilities in SNe and neutron-star mergers, it was often taken for granted that they rapidly grow nonlinear, leaving the question of seeding largely unexamined. Our estimates suggest, however, that seeding needs to be addressed quantitatively for neutrino-mass-induced instabilities, which are the first to appear in SNe.
Our framework sheds new conceptual light on this question. Neutrino phase-space density fluctuations directly seed unstable modes, as per Eq.~\eqref{eq:seeding}, whereas matter inhomogeneities cannot directly seed instabilities. Yet the forces they exert
on flavomons can transfer amplitude from larger to smaller scales, potentially feeding instability growth.

For neutrino-mass-induced instabilities, the first to appear in SNe, matter gradients may limit growth before saturation, making seeding and amplitude evolution crucial ingredients. Our amplitude-transport equation~\eqref{eq:transport_flavomons} incorporates both within a linear ray-tracing framework. Our result provide a crucial starting point for understanding the global role of flavor conversion in SNe.

{\bf\textit{Note added}}---While this work was finalized, Ref.~\cite{Zaizen:2026wvj} appeared, studying fast instabilities in a global matter profile through numerical solutions of the quantum kinetic equations with attenuated neutrino--neutrino refraction. Applying the two-beam criterion of Ref.~\cite{Bhattacharyya:2025gds} to a continuum angular distribution, they obtain an adiabaticity condition based on the picture that the growing mode turns stable after $\Delta t\sim \gamma/(\partial \Omega_R/\partial t)$, where $\Omega_R$ is the real part of the eigenfrequency. On the other hand, our first-principles WKB approach reveals a timescale $\Delta t\sim \Delta K/(\partial \Omega_R/\partial r)$. There may be no contradiction in the fast case, where the two scales could be comparable, but for slow instabilities, the focus of our work, they are completely different. We have also obtained a specific criterion, Eq.~\eqref{eq:matter_gradient_criterion_fast}, for weak fast instabilities, which is parametric and does not depend on numerically solving the dispersion relation. We agree that attenuation can artificially suppress instability growth. Our focus, instead, is on slow instabilities, which may be suppressed below the shock wave even in the absence of such attenuation. The general problem of global evolution can be described by our new Eq.~\eqref{eq:transport_flavomons}, valid in the linear regime without additional approximations or artificial attenuation.

{\bf\textit{Acknowledgments}}---We thank Luke Johns for helpful comments on the manuscript. DFGF thanks the organizers and participants of the workshop ``Collective Neutrino Oscillations in Supernovae and Neutron Star Mergers,'' and especially Sajad Abbar, Soumya Bhattacharyya, Hiroki Nagakura, Meng-Ru Wu, Zewei Xiong, for many fruitful conversations. DFGF acknowledges support by the TAsP (Theoretical Astroparticle Physics) project. GGR acknowledges partial support by the German Research Foundation (DFG) through the Collaborative Research Centre ``Neutrinos and Dark Matter in Astro- and Particle Physics (NDM),'' Grant SFB--1258--283604770, and under Germany’s Excellence Strategy through the Cluster of Excellence ORIGINS EXC--2094--390783311.

\bibliographystyle{bibi}
\bibliography{References}

\onecolumngrid

\clearpage

\begin{center}
\textbf{\large End Matter}
\end{center}
\bigskip
\twocolumngrid

{\bf\textit{Seeding by flavor mixing}}---The growth of unstable flavor waves requires initial disturbances. They must be provided by the neutrino mass matrix, which is the only flavor-violating ingredient in our system. If all neutrinos begin in flavor eigenstates, then in the quantum kinetic equation~\eqref{eq:QKE}, written in the weak interaction basis, the only off-diagonal piece comes from $\sH^{\rm vac}_\sp$ and thus from $s_V=\sin2\theta_V$. To understand the resulting onset dynamics for the instability growth, our starting point consists of the exact equations for the field of flavor coherence,
\begin{eqnarray}\label{eq:EoM-psi}
    v^\mu \partial_\mu\psi_\sp&=&i(\lambda^\mu+D^\mu)v_\mu \psi_\sp-i\psi^\mu v_\mu D_\sp
    \nonumber\\
    &&\kern3em{}-i\omega_E\left(c_V\psi_\sp+ s_V D_\sp\right),
\end{eqnarray}
where we recall the phase-space notation $\sp=\{E,\bv\}$ for ultrarelativistic neutrinos with $-\infty<E<+\infty$. In this End Matter we set $\mu=1$ for compact notation; $\mu$ can indeed always be absorbed in the definition of $D_\sp$.

Physically, the mixing term seeds the unstable eigenmode, which then grows through mixing-free collective dynamics. It is not, however, the homogeneous neutrino phase-space density $D_{\sp}$ that directly seeds small-scale modes. To describe seeding, we first introduce scale separation through $D_\sp\to D_\sp+\delta D_\sp$, where $D_\sp$ denotes the slowly varying background and $\delta D_\sp$ its small-scale fluctuations with $\delta D_\sp\ll D_\sp$. For now, we take the background to be homogeneous, retaining fluctuations on scales of order $|\bK|^{-1}$.

We now need the projection of the mixing seed onto the unstable eigenmode. To find it, we turn to the one-sided Fourier transform that we also used in Ref.~\cite{Fiorillo:2026byh}, analogous to the Laplace transform in Ref.~\cite{Fiorillo:2024bzm},
\begin{equation}
    \psi_{\sp,K}=\int_0^{\infty} \! dt \int d^3\br\,\psi_\sp e^{i K\cdot x},
\end{equation}
where $K=(\Omega,\bK)$ and $K\cdot x=\Omega t-\bK\cdot\br$. We thus find
\edit{\begin{equation}
    \psi_{\sp,K}=\frac{i\psi_{\sp,\bK}(0)+(\psi_K^\mu v_\mu D_\sp+i\omega_E s_V/\Omega\delta D_{\sp, \bK})   
    }{k\cdot v-\omega_E c_V}.
\end{equation}
Here, $\psi^\mu=\sum_\sp \psi_\sp v^\mu$, while $\psi_{\sp,\bK}(t)$ and $\delta D_{\sp,\bK}$ denote the spatial Fourier components of flavor coherence and density fluctuations respectively.}

Multiplying by $v^\mu$ and integrating over neutrino phase space yields
\begin{equation}
\varepsilon_{\mu\nu}\psi_K^\nu=i\sum_\sp
\frac{\psi_{\sp,\bK}(0) v^\mu}{k\cdot v-\omega_E c_V}+i\sum_\sp\frac{\omega_E s_V \delta D_{\sp,\bK} v^\mu}{\Omega(k\cdot v-\omega_E c_V)}.    
\end{equation}
The unstable eigenmode emerges as a zero-eigenvector of the dielectric function $\varepsilon_{\mu\nu} n^\nu=0$ for a certain complex frequency $\Omega=\Omega_R+i\gamma$. Formally, we can separate a contribution $\varepsilon^{\mu\nu}_{\rm pole}=\mathcal{R}(\Omega-\Omega_R-i\gamma) n^\mu n^\nu$, where we normalize the eigenvector $n^\mu$ such that $n^0=1$.

For $\Omega$ close to the eigenmode frequency, we then express the solution $\psi^\mu_K$ as the superposition of a contribution from the initial condition $\psi_{p,\bK}(0)$ and from the seeding term. Considering only the latter, which acts as a source for the unstable eigenmode, we find
\begin{equation}
    \psi^\mu_K=\frac{i n^\mu}{\mathcal{R}n^4 \Omega (\Omega-\Omega_R-i\gamma)}\sum_\sp\frac{\omega_E s_V \delta D_{\sp,\bK} v\cdot n}{k\cdot v-\omega_E c_V},
\end{equation}
where $n^4=(n_\mu n^\mu)^2$. We can now perform the inverse Fourier transform in time to obtain the exponentially growing contribution, in the same way as in Refs.~\cite{Fiorillo:2026byh, Fiorillo:2024bzm}
\begin{equation}
    \psi^\mu_\bK=n^\mu\,\frac{e^{-i(\Omega_R+i\gamma)t}}{(\Omega_R+i\gamma)\partial_\Omega \varepsilon_{\alpha\beta}n^\alpha n^\beta}\sum_\sp\frac{\omega_E s_V \delta D_{\sp,\bK} v\cdot n}{k\cdot v-\omega_E c_V},
\end{equation}
where now the expression is evaluated at $\Omega=\Omega_R+i\gamma$. We have used the identity $\mathcal{R} n^4=\partial_\Omega \varepsilon_{\alpha\beta}n^\alpha n^\beta$. If we take the time derivative of this term, and focus on $\mu=0$, we find that the amplitude of the unstable mode $\psi^0_\bK$ satisfies the equation
\begin{eqnarray}
    \partial_t\psi^0_\bK&=&-i(\Omega_R+i\gamma)\psi^0_\bK
    \nonumber\\[1.0ex]
    &&{}-\frac{i}{\partial_\Omega \varepsilon_{\alpha\beta}n^\alpha n^\beta}\sum_\sp\frac{\omega_E s_V \delta D_{\sp,\bK} v\cdot n}{k\cdot v-\omega_E c_V}.
\end{eqnarray}
The second line is the seeding term reported in Eq.~\eqref{eq:seeding} of the main text. Coupled with the WKB drift terms, these provide a description of the unstable wavepackets in a weakly inhomogeneous. \edit{We recognize in the mixing term the direct projection of the seeding term $-i\omega_E s_V \delta D_{\sp,\bK}$ onto the unstable eigenvector, multiplied by the wavefunction renormalization $\sqrt{\mathcal{Z}}$.
}

\edit{The seeding term was derived in a large-scale homogeneous environment. Relaxing this assumption, $\delta D_\sp$ varies both on large scales, reflecting the background inhomogeneity, and on the small scales already included. This scale separation allows us to introduce a local Fourier component $\delta D_{\sp,\bK}(\br)$, depending on both $\br$ and $\bK$. We define it by a Fourier transform within a ball centered at $\br$ with radius $\ell\gg |\bK|^{-1}$ much smaller than the inhomogeneity scale and volume $V_\ell=4\pi \ell^3/3$
\begin{equation}\label{eq:inverse_transform}
    \delta D_{\sp,\bK}=V_\ell^{-1}\int_{V_\ell} \delta D_{\sp}(\br+\br') e^{-i\bK\cdot \br'}dV'.
\end{equation}
Scale separation makes the precise value of $\ell$ irrelevant. This local transformation connects the seeding term to the inhomogeneous environment, where $\Psi$ evolves along the rays. Indeed, applying the same transformation to $\Psi$, we find
\begin{equation}
    \frac{1}{V_{\ell}}\int_{V_\ell} \mathcal{A}(\br+\br') e^{i S(\br)+i \boldsymbol{\nabla}S \cdot \br'} e^{-i\bK\cdot \br'}dV'=\mathcal{A}(\br) e^{iS(\br)}
\end{equation}
to leading WKB order, provided $\bK=\nabla S(\br)$ is the local wavevector.
}

\edit{The formalism also clarifies the possible role of matter-density fluctuations in seeding flavor instabilities. Unlike neutrino phase-space density fluctuations, they cannot act as a direct seed: in Eq.~\eqref{eq:EoM-psi}, matter fluctuations couple to $\psi_\sp$ and therefore cannot generate flavor coherence when $\psi_\sp=0$. Instead, small-scale matter fluctuations $\delta\lambda^\mu$ can transfer existing flavor coherence from large to small scales. In the kinetic equation for $\psi_\sp$, they produce a term of the form
\begin{equation}\label{eq:source_matter_small_scales}
    (\partial_t \psi_{\sp,\bK})_{\delta \lambda}=i\sum_{\bf q} \delta \lambda^\mu_{\bf q}v_\mu \psi_{\sp,\mathbf{K}-\mathbf{q}}.
\end{equation}
Thus, matter fluctuations can seed small-scale modes only indirectly, by transferring pre-existing flavor coherence from larger scales. In the limit $q\ll K$, this mode coupling is described by the continuous WKB wavevector drift $d\bK/dt$ derived in the main text.


In conclusion, the question of seeding has long been swept under the carpet, with most numerical simulations assuming small-scale neutrino phase-space fluctuations whose physical origin is uncertain.  We have shown that large-scale density fluctuations can instead provide seeds through mixing-induced flavor coherence, which is then transferred to small-scale unstable modes. This transfer can occur through the WKB wavevector drift caused by matter gradients, or through coupling to small-scale matter density fluctuations not captured by the WKB method. 
}

\onecolumngrid

\onecolumngrid

\setcounter{equation}{0}
\setcounter{figure}{0}
\setcounter{table}{0}
\setcounter{page}{1}
\makeatletter
\renewcommand{\theequation}{S\arabic{equation}}
\renewcommand{\thefigure}{S\arabic{figure}}
\renewcommand{\thepage}{S\arabic{page}}

\begin{center}
\textbf{\large Supplemental Material for the Letter\\[0.5ex]
\mbox{Flavomons in Matter Gradients: Ray Tracing and Amplitude Evolution}}
\end{center}
\bigskip

\twocolumngrid

In this Supplemental Material, we provide some details on the pinched quasi-thermal spectrum used for the example of Fig.~1 in the main text and on the resulting dispersion relation for slow modes. Moreover, we provide a formal derivation of the amplitude transport equation~\eqref{eq:transport_flavomons}. We further show formally, that gravitational effects are subdominant for flavomon propagation, in contrast to the underlying neutrinos.

\begin{table}[b!]
\caption{Parameters for the pinched distribution used for the schematic representation of neutrinos below the shock wave.
\label{tab:neutrino_parameters}}
\vskip2pt
\begin{tabular*}{\columnwidth}{@{\extracolsep{\fill}} llllll}
\toprule
Species & $\mathcal{N}_i$ [$\times 10^{34}\,\mathrm{cm}^{-3}$] & $\overline{E}_i$ [MeV] & $\alpha_i$ & $\delta \theta_i^2$ & $\delta_i$ \\
\midrule
$\nu_e$           & 2.12 & 9.0 &  3.3 & 0.08  & 0.02  \\
$\overline{\nu}_e$ &  2.66 & 12.2 & 3.5  & 0.05  &  0 \\
$\nu_x$            &  0.82 &   12.3 & 2.5  & 0.035 & 0 \\
\bottomrule
\end{tabular*}
\end{table}

{\bf\textit{Schematic neutrino distribution}}---To examine the impact of matter gradients on the evolution of slow instabilities, we use a schematic model for the neutrino energy and angular distribution. Specifically, we adopt for each species $i=\nu_e$, $\overline{\nu}_e$, or $\nu_x$ a pinched quasi-thermal spectrum of the form
\begin{equation}
    \frac{dN_{i}}{dEd\cos\theta}=\frac{d\mathcal{N}_i}{d\cos\theta}\frac{1}{\overline{E}_i}\left(\frac{E}{\overline{E}_i}\right)^{\alpha_i}e^{-(1+\alpha_i)E/\overline{E}_i}
\end{equation}
where $\overline{E}_i$ is the average energy and $\alpha_i$ the pinch parameter. We model the angular dependence as
\begin{equation}
    \frac{d\mathcal{N}_i}{d\cos\theta}=\mathcal{N}_i\left[e^{-(1-\cos\theta)^2/\delta\theta_i^4}+\delta_i\right],
\end{equation}
where $\delta\theta_i$ measures the typical angular range of the given species around the forward direction. Moreover, $\delta_i$ represents a weak backward flux, notably for $\nu_e$, which is the most strongly coupled species. 

\begin{figure}
    \centering
    \includegraphics[width=\columnwidth]{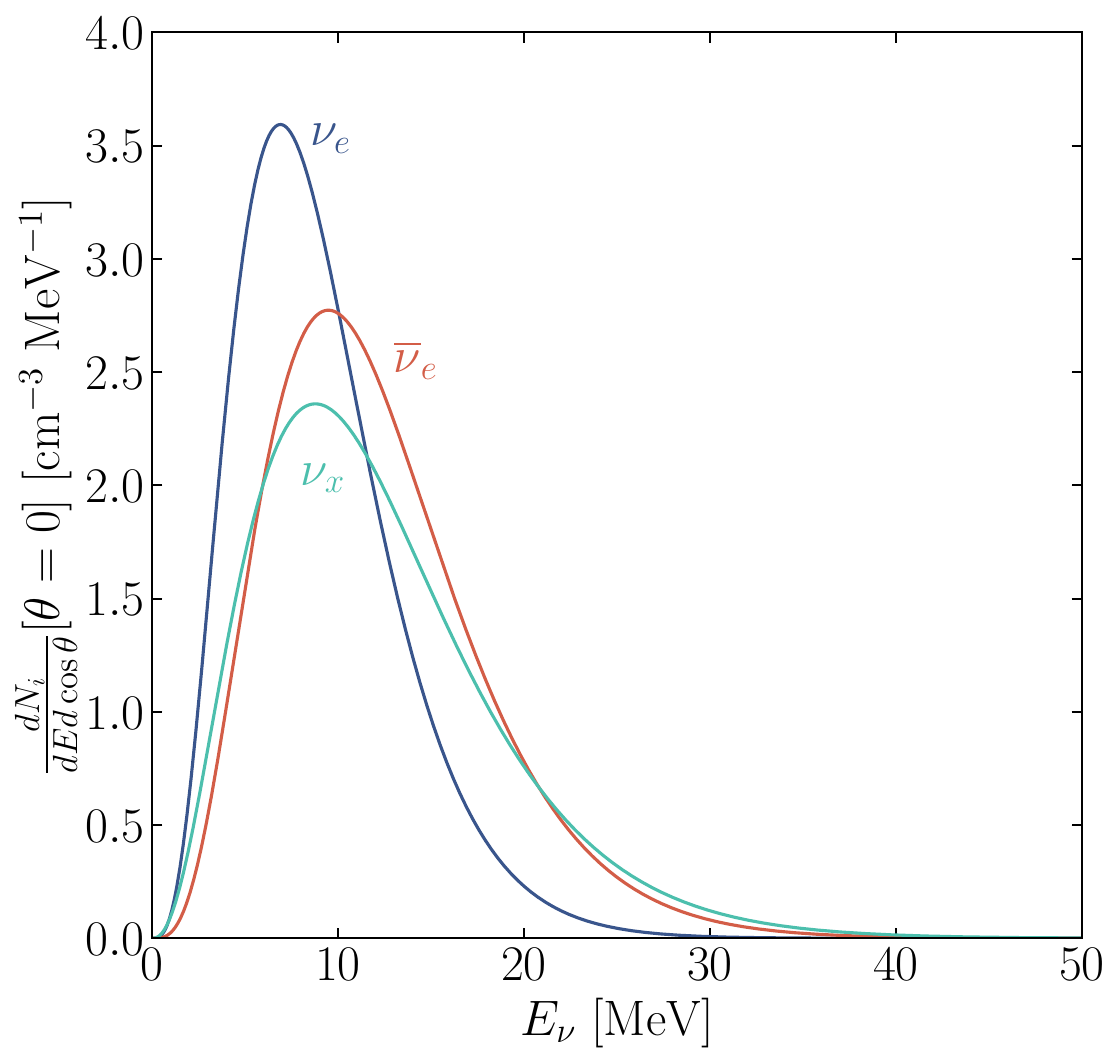}
    \caption{Energy spectra of neutrinos moving radially ($\theta=0$). The species $\nu_x$ represents both $
    \nu_\mu$ and $\overline{\nu}_\mu$, taken to have equal distributions.}
    \label{fig:spectra}
\end{figure}

The parameter values to reproduce the distribution in Fig.~1 of the main text are collected in Table~\ref{tab:neutrino_parameters}. The total number density of neutrinos, integrated over energy and direction, follows the expected hierarchy, with $\nu_e$ dominating and $\nu_x$ the least populated species. The average energies show the opposite ordering, with $\nu_x$ being more energetic since they come from deeper regions. The pinch parameters are roughly representative of the typical values inferred, e.g., from the simulations shown in Ref.~\cite{Serpico:2011ir}.  The spectra for radially moving neutrinos are shown in Fig.~\ref{fig:spectra}, which clearly reveals that the $\nu_e$ are more numerous, whereas  and the $\nu_x$ have a harder spectrum. The width of the opening angle $\delta\theta_i$ is narrower for $\nu_x$, which decouple at smaller radii and therefore are more strongly forward focused.


{\bf\textit{Dispersion relation of slow modes}}---We briefly review the main properties of slow instabilities, starting from the dielectric tensor $\varepsilon^{\mu\nu}$ of Eq.~\eqref{eq:dielectric_tensor}. In our context, the main characteristic is that, if $\omega_E\ll \mu \epsilon^2$, the unstable modes are very close to the light cone~\cite{Fiorillo:2024pns, Fiorillo:2025ank, Fiorillo:2025zio}. Therefore, we can choose $\omega_\bk=\pm|\bk|+\chi$, with $\chi\sim \omega_E\ll \mu\epsilon$. For these near-luminal modes, the denominator $k\cdot v-\omega_E c_V=|\bk|(\pm1-\bkappa\cdot\bv)+(\chi-\omega_E c_V)$ nearly vanishes, where $\bkappa=\bk/|\bk|$. Therefore, we can separate the dielectric tensor in a logarithmic and a regular contribution
\begin{equation}\label{eq:approx_dielectric}
    \varepsilon^{\mu\nu}=g^{\mu\nu}-\kappa^\mu \kappa^\nu\varepsilon_{\rm log}-d^{\mu\nu},
\end{equation}
where $\kappa^\mu=(1,\bkappa)$. The logarithmic contribution obtains from evaluating the numerator exactly near the singularity $v^\mu=\kappa^\nu$, so that
\begin{equation}
    \varepsilon_{\rm log}=\sum_\sp \frac{\mu D_{\sp}\vert_{\bv=\bkappa}}{|\bk|(\pm1-\bkappa\cdot \bv)+\chi-\omega_E c_V},
\end{equation}
where $D_{\sp}\vert_{\bv=\bkappa}$ is the DLN evaluated in the direction of the mode. 

The regular contribution obtains by subtracting the DLN in this direction, removing the singularity at $\bv=\bkappa$. Therefore, in the denominator we may neglect $\chi$ and $\omega_E$, implying
\begin{equation}
    d^{\mu\nu}=\sum_\sp\frac{\mu(v^\mu v^\nu D_\sp-\kappa^\mu \kappa^\nu D_{\sp}\vert_{\bv=\bkappa})}{|\bk|(\pm1-\bkappa\cdot\bv)}.
\end{equation}
Notice that $d^{\mu\nu}$ does not depend on the frequency of the mode.  

The logarithmic contribution can be made more \hbox{explicit} by integrating over directions. To this end, we separate $\sum_\sp=\int E^2 dE\,d^2\bv/(2\pi)^3$ and introduce the energy distribution $D_\bkappa(E)=D_{\sp}\vert_{\bv=\bkappa} E^2/(2\pi)^3$. The integral over \hbox{velocities} in $\varepsilon_{\rm log}$ can now be done directly, as in Refs.~\cite{Fiorillo:2024pns,Fiorillo:2025zio}, and gives
\begin{equation}
    \varepsilon_{\rm log}=\pm \int dE\,\frac{2\pi\mu D_{\bkappa}(E)}{|\bk|}\log\left[\frac{\pm 2|\bk|}{\chi-\omega_E c_V}\right],
\end{equation}
where $\log(\pm 2|\bk|)$ acquires a $+i\pi$ contribution if we choose the minus sign, and $\log(\chi-\omega_E)=\log|\chi-\omega_E|+i\phi$ with $-\pi/2<\phi<3\pi/2$; the choice of branches of the logarithm is directly determined by the retarded prescription of integration~\cite{Fiorillo:2023mze,Fiorillo:2024bzm,Fiorillo:2024uki,Fiorillo:2024pns,Fiorillo:2025zio}.

The growth rate of an instability moving along an arbitrary direction $\bkappa$ can be determined already at this stage. The dispersion relation is $\mathrm{det}[\varepsilon^\mu_\nu]=0$, giving an equation for $\chi$ as a function of $|\bk|$ at a fixed $\bkappa$. There are multiple solutions, corresponding to different branches. For modes with a polarization vector $\psi^\mu$ such that \hbox{$\kappa\cdot\psi=0$}, the logarithmic contribution drops out; there remains a logarithmic term for these modes, but it requires continuing the expansion for small $\chi-\omega_E c_V$ to the next order, as done in Ref.~\cite{Fiorillo:2024pns} for the axial-breaking modes. 

Let us instead consider the modes whose polarization vector is \textit{not} orthogonal to $\kappa^\mu$. In that case, for the determinant equation to vanish, we require $\mathrm{Im}(\varepsilon_{\rm log})=0$, since in that equation $\varepsilon_{\rm log}$ is the only complex term, and the dependence on $\chi$ is entirely through it. Let us focus on the branch with $\omega_\bk=-|\bk|+\chi$; we showed in Refs.~\cite{Fiorillo:2024pns,Fiorillo:2025zio} that in normal ordering this is the branch that becomes unstable for axially symmetric distributions. We choose $\chi=\chi_R+i\gamma$, and require $\mathrm{Im}(\varepsilon_{\rm log})=0$, so
\begin{equation}\label{eq:dispersion_slow}
    \int dE\left[D_{\bkappa}(E)\left(\arctan\left(\frac{\gamma}{\chi_R-\omega_E c_V}\right)-\pi\right)\right]=0.
\end{equation}
The arctan obtains with the same prescription as the logarithm, i.e., it runs from $-\pi/2$ to $3\pi/2$. This equation is independent of $|\bk|$, and directly provides $\gamma(\chi_R)$, the growth rate as a function of the separation from the light cone. For very small $\gamma$, it receives contributions only from neutrinos with $\omega_E <\chi_R/c_V$, corresponding to the kinematic condition for flavomon emission that was discussed in the main text and in Ref.~\cite{Fiorillo:2025kko}. 

Equation~\eqref{eq:dispersion_slow} provides a practical method to determine the growth rate of instabilities with $|\chi|\ll\mu \epsilon$, while it may fail if the DLN vanishes across some directions, in which case the instability becomes nonresonant with a stronger growth rate. For this case, a complementary approximation method is developed in Ref.~\cite{Fiorillo:2025zio}.

\begin{figure}[b]
    \centering
    \includegraphics[width=\columnwidth]{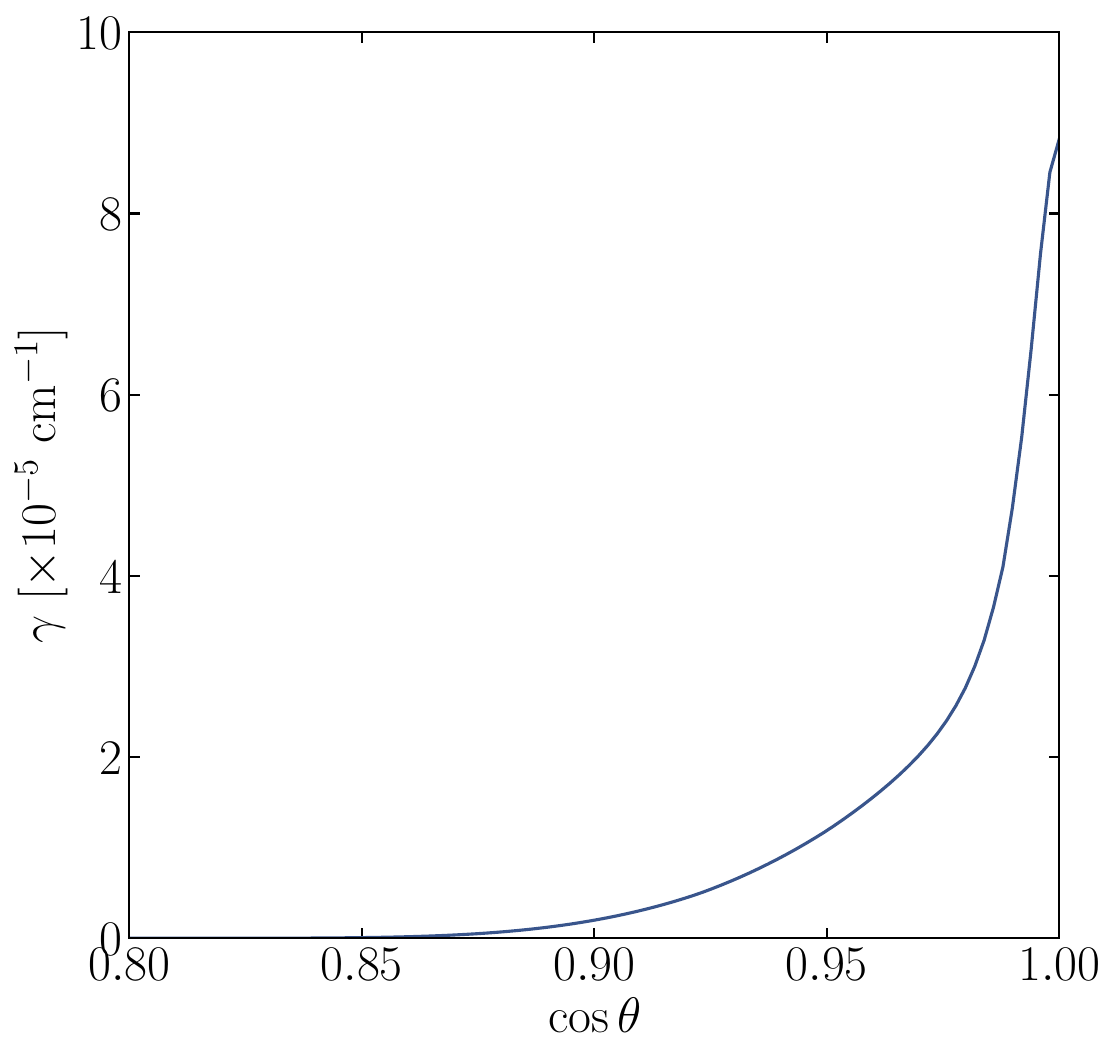}
    \caption{Maximum growth rate of unstable modes as a function of the mode direction.}
    \label{fig:growth_rate_direction}
\end{figure}

We now use this equation to deduce the maximum growth rate for the schematic neutrino distributions introduced in the previous section and used in the main text. The DLN distribution is directly determined by the neutrino ones through
\begin{equation}
    D_\bkappa(E)=\frac{1}{2\pi}
    \begin{cases}
    \frac{dN_{\nu_e}}{dEd\cos\theta}-\frac{dN_{\nu_x}}{dEd\cos\theta},\;E>0,\\[2ex]
    \frac{dN_{\overline{\nu}_x}}{dEd\cos\theta}-\frac{dN_{\overline{\nu}_e}}{dEd\cos\theta},\;E<0.
    \end{cases}
\end{equation}
We use the largest mass splitting in $\omega_E=\delta m^2/2E$ with $\theta_V=0$ and $\delta m^2=2.5\times 10^{-3}\,\mathrm{eV}^2$, corresponding to normal ordering. With this choice, a neutrino with typical energy $\overline{E}_i\sim 10$\,MeV has a vacuum frequency $\omega_E\sim10^{-5}\,\mathrm{cm}^{-1}$.

The resulting growth rate is shown in Fig.~\ref{fig:growth_rate_direction}. The strong peak for forward directions, similar to what was found in the analysis of Ref.~\cite{Fiorillo:2025gkw}, is due to the stronger contribution of $\overline{\nu}_e$ and high-energy $\nu_\mu$, which are flipped compared to the dominant DLN. The growth rate itself turns out to be of the order of $\gamma\sim 10^{-5}$--$10^{-4}\,\mathrm{cm}^{-1}$. This is completely consistent with the estimates in the main text, since the typical asymmetry in the global DLN distribution is $\epsilon\sim 0.5$, while along the forward direction, it can be smaller due to the larger fraction of $\overline{\nu}_e$ moving forward, so that $\omega_E/\epsilon\sim 10^{-5}$--$10^{-4}$.

To relate the unstable modes to their wavenumber, we have to return to Eq.~\eqref{eq:approx_dielectric}; solving it generally requires diagonalizing the matrix $\varepsilon^{\mu\nu}$. However, since the dominant instability is in the forward direction ($\bk=k_z \hat{\bf{z}}$), we may obtain $\gamma(|\bk|)$ much more easily using the axial symmetry of the distribution. In particular, as shown in Ref.~\cite{Fiorillo:2025zio}, the dispersion relation of longitudinal modes moving along the axis of symmetry with $\omega=k_z+\chi$ is
\begin{eqnarray}\label{eq:dispersion_kz}
    \kern-1em&&2\pi\int dE D_\bkappa(E) \log\left[\frac{2k_z}{\chi-\omega_E}\right]={}
    \\[1ex] \nonumber
    \kern-1em&&\frac{\left(\frac{k_z}{\mu}\right)^2-\left(\frac{k_z}{\mu}\right)(D_0+D_1)+D_0^2+d_0(D_1-D_0)}{D_0-D_1},
\end{eqnarray}
where $D_0=\sum_\sp D_\sp$ and $D_1=\sum_\bp D_\bp \cos \theta$, whereas
\begin{equation}
    d_0=\sum_\sp \frac{D_\sp-D_\sp\vert_{\bv=\bkappa}}{1-\cos\theta},
\end{equation}
with $\cos\theta$ the angle between $\bv$ and the symmetry axis.

\begin{figure}
    \centering
    \includegraphics[width=\columnwidth]{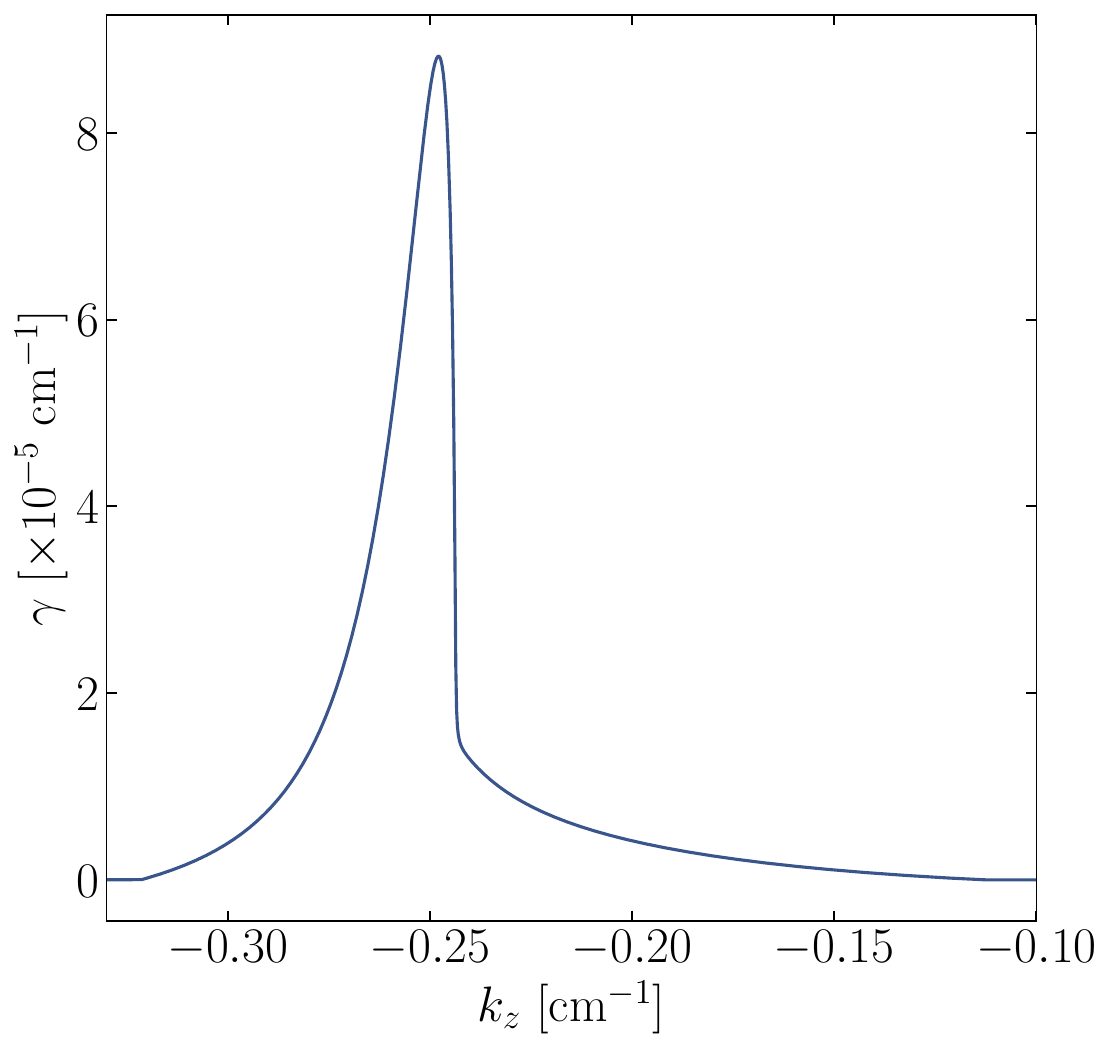}
    \caption{Growth rate for forward-moving modes, as a function of the wavenumber $k_z$.}
    \label{fig:growth_k}
\end{figure}

Solving Eq.~\eqref{eq:dispersion_kz} gives the growth rate for the forward-moving modes, shown in Fig.~\ref{fig:growth_k}. As already derived in Refs.~\cite{Fiorillo:2024pns, Fiorillo:2025ank, Fiorillo:2025zio}, for $\omega\simeq k_z$ in normal ordering, the instability appears for negative $k_z$. Notice that the approximate dispersion relation Eq.~\eqref{eq:dispersion_kz} assumes that the angular distribution changes slowly. Specifically, since we replaced $|\bk|(\pm 1-\bkappa\cdot\bv)+\chi-\omega_E c_V\simeq|\bk|(\pm 1-\bkappa\cdot\bv)$, we require that the distribution changes slowly over angles $\delta\theta^2\sim |\chi|/|\bk|$. Since our benchmark distribution changes over angles $\delta \theta^2\sim 10^{-2}$, while $|\chi|\sim 10^{-4}\,\mathrm{cm}^{-1}$, this means that our approximation only holds for $|k_z|\gg 10^{-2}\,\mathrm{cm}^{-1}$.

Most importantly, even though the growth rate has a typical magnitude $\gamma\sim 10^{-4}\,\mathrm{cm}^{-1}$, the interval of unstable wavenumber is much larger, as it is determined by $\mu \epsilon$; indeed, the unstable $K$ range is roughly $\Delta k=\Delta K\sim 0.1\,\mathrm{cm}^{-1}$. In particular, the maximum number of e-folds accumulated within a matter gradient $\partial_r\lambda$ can be directly determined from the flavomon equations of motion~as
\begin{equation}
    N_{\text{e-folds}}=\frac{\int \gamma dK}{|\partial_r\lambda|}=\frac{2.4\times 10^{-6}\,\mathrm{cm}^{-2}}{|\partial_r\lambda|},
\end{equation}
confirming the order-of-magnitude estimate shown in the main text.

{\bf\textit{Amplitude evolution along rays}}---Equation~\eqref{eq:transport_flavomons} of the main text describes the propagation of flavomon amplitudes along geometrical-optics rays. At this level, flavor waves are simply scalar waves propagating in a slowly varying dispersive medium, so Eq.~\eqref{eq:transport_flavomons} is the standard WKB transport equation supplemented by the source term derived in the End Matter. In particular, the drift follows the group velocity, while the term proportional to $\boldsymbol{\nabla}\cdot \bv_{\rm gr}$ accounts for geometrical focusing and ensures conservation of wave action in the absence of sources.

While in the main text, this result was obtained somewhat heuristically from geometrical optics, flavor waves emerge microscopically as collective excitations of the neutrino flavor-coherence fields $\psi_\sp(x,t)$. It is therefore instructive to derive Eq.~\eqref{eq:transport_flavomons} directly from the underlying neutrino kinetic equations. This derivation identifies the appropriate normalization of the flavomon amplitude, including the wavefunction renormalization factor $\mathcal{Z}$, and shows explicitly how the standard geometrical-optics transport law emerges from the microscopic dynamics.

\edit{
Once again, we start from the linearized equations for the fields of flavor coherence, $\psi_{\sp}(x,t)$, where we recall the phase-space notation $\sp=\{E,\bv\}$ for ultrarelativistic neutrinos with $-\infty<E<+\infty$. Compared with Eq.~\eqref{eq:EoM-psi}, we neglect the seeding term proportional to $s_V$ and~use
\begin{equation}
    v^\mu \partial_\mu \psi_{\sp}=i(\lambda^\mu + D^\mu) v_\mu \psi_{\sp}-i\psi^\mu v_\mu D_{\sp} -i\omega_E c_V \psi_{\sp}.
\end{equation}
As in the End Matter, we set $\mu=1$ for compact notation. The usual derivation of flavomon dispersion begins with the ansatz \smash{$\psi_{\sp}(x)=\tilde{\psi}_{\sp} e^{-i K\cdot x}$}, where the amplitudes \smash{$\tilde{\psi}_{\sp}$} are found as the eigenfunctions of collective modes. In a slowly varying medium, instead we begin with the eikonal ansatz $\psi_{\sp}(x)=\tilde{\psi}_{\sp}(x) e^{i S(x)}$, where the eikonal $S(x)$ varies rapidly in space and time, \smash{$\tilde{\psi}_{\sp}(x)$} are slowly varying amplitudes, and $\partial_\mu S(x)=- K^\mu(x)$ defines the local wave vector. Notice that $\partial_t S$ corresponds only to the real part of the eigenfrequency; the imaginary part, assumed to be very small for weakly unstable modes, affects the amplitude, not the eikonal. The leading terms containing only the derivatives of $S$ on the left-hand side reproduce the usual dispersion relation, which we write in the form $\mathrm{Re}(\varepsilon_{\mu\nu})(K,x)\, n^\nu=0$; for weakly unstable modes, the imaginary part of the dielectric function can be neglected to first approximation~\cite{Fiorillo:2024bzm,Fiorillo:2024uki}; its contribution simply leads to the growth of the amplitude with a rate $\gamma_\bK$, as shown in Ref.~\cite{Fiorillo:2024uki}. Hereafter, we simply write $\varepsilon_{\mu\nu}$ to denote the real part, since the imaginary part is by assumption small.
}

The subleading terms describing slow amplitude variation are obtained by keeping the derivatives of~$\tilde{\psi}_{\sp}$
\begin{equation}
    q_\sp\tilde{\psi}_{\sp}=D_\sp v\cdot \tilde{\psi}-i v\cdot \partial \tilde{\psi}_\sp,
\end{equation}
where $q_\sp=(K+\lambda+D)\cdot v-\omega_E c_V$ and $\tilde{\psi}^\mu=\sum_\sp\tilde{\psi}_\sp v^\mu$. We can proceed iteratively to solve for \smash{$\tilde{\psi}_\sp$}, which gives
\begin{equation}
    \varepsilon^{\mu\nu}\tilde{\psi}_\nu=-i\sum_\sp\frac{v^\mu v^\alpha}{q_\sp}\partial_\alpha \left(\frac{D_\sp v\cdot\tilde{\psi}}{q_\sp}\right).
\end{equation}
We choose the field satisfying the local dispersion relation as $\tilde{\psi}^\mu=An^\mu+\tilde{\psi}^1_\mu$, where $\tilde{\psi}^1_\mu$ is a small correction that can be chosen orthogonal to $n^\mu$; since $\varepsilon^{\mu\nu} n_\nu$, when we multiply the equation by $n^\mu$ we find
\begin{equation}
    \sum_\texttt{p}\frac{v\cdot n }{q_\sp}(v\cdot \partial)\left(A\frac{D_\sp v\cdot n}{q_\sp}\right)=0.
\end{equation}
Introducing the vector $\mathcal{J}^\alpha=\sum_\sp (v\cdot n)^2 v^\alpha D_\sp/q_\sp^2$, we may rewrite this expression as
\begin{equation}
    \mathcal{J}^\alpha \partial_\alpha A+ A\sum_\sp\frac{v\cdot n}{q_\sp} (v\cdot\partial)\left(\frac{D_\sp v\cdot n}{q_\sp}\right)=0.
\end{equation}
After rearranging the last term, we find
\begin{equation}
    \mathcal{J}^\alpha \partial_\alpha A+\frac{A}{2}\partial_\alpha \mathcal{J}^\alpha+\frac{A}{2}\sum_\sp\frac{(v\cdot n)^2}{q_\sp^2}v\cdot \partial D_\sp=0.
\end{equation}
For a collisionless neutrino plasma, $v\cdot \partial D_\sp=0$, and since we neglect collisions with matter, we drop this term. 

The second term has instead a very simple interpretation: we can rewrite the first two terms together as
\begin{equation}
    \partial_\alpha\left(|A|^2 \mathcal{J}^\alpha\right)=0.
\end{equation}
Here we have used the assumption that $\mathcal{J}^\alpha$ is a real quantity, which is approximately true for weakly unstable or weakly damped states.
The vector $\mathcal{J}^\alpha$ is equal to
\begin{equation}
    \mathcal{J}^\alpha=n^\mu n^\nu \frac{\partial \varepsilon_{\mu\nu}}{\partial K_\alpha};
\end{equation}
if we differentiate the equation $\varepsilon_{\mu\nu}[\Omega(\bK),\bK]n^\nu(\bK)=0$ with respect to $\bK$, and multiply by $n^\mu(\bK)$, we find the identity $\mathcal{J}^i=v_{\rm gr}^i\mathcal{J}^0$, where $\bv_{\rm gr}=\partial\Omega/\partial \bK$ is the group velocity. 

Furthermore, noting that $\mathcal{J}^0$ can be rewritten as $\mathcal{J}^0=n^\mu n^\nu \partial_\Omega\varepsilon_{\mu\nu}=\mathcal{Z}^{-1}$, where $\mathcal{Z}$ is the wavefunction renormalization introduced in Ref.~\cite{Fiorillo:2025npi}, we finally find
\begin{equation}
    \partial_\alpha\left(\mathcal{Z}^{-1} |A|^2 U^\alpha\right)=0,
\end{equation}
where $U^\alpha=(1,\bv_{\rm gr})$. Recognizing $|A|^2 \mathcal{Z}^{-1}$ as flavomon number density for a near-stable branch~\cite{Fiorillo:2025npi}, we understand this as the equation of continuity for the flavomon number, sometimes called the wave action conservation law. Notice that the normalization choice of $n^\mu$ is, in principle, arbitrary, but the arbitrariness disappears in the physical quantity $A/\sqrt{\mathcal{Z}}$ as it should. A convenient choice is $n^0=1$, so that $A$ coincides with the amplitude of the field $A=\tilde{\psi}^0_\bK$: we add the suffix $\bK$ to indicate that only this specific wave vector was selected by the eikonal solution. The physical field describing the amplitude of the flavomon mode is $\Psi_\bK=\tilde{\psi}^0_\bK e^{i S}/\sqrt{\mathcal{Z}}$, incorporating the eikonal factor.

If we combine this last equation with the seeding term derived in the End Matter, which is possible due to the linear nature of the equations, we finally find
\begin{eqnarray}
   \kern-3em &&(U\cdot \partial)\Psi_\bK+\frac{1}{2}\Psi_\bK \boldsymbol{\nabla}\cdot \bv_{\rm gr}
    \nonumber\\
    \kern-3em&&{}=(\gamma_\bK-iU\cdot K) \Psi_\bK-i\sqrt{\mathcal{Z}}\sum_\sp\frac{\omega_E s_V \delta D_{\sp,\bK} v\cdot n}{k\cdot v-\omega_Ec_V};
\end{eqnarray}
\edit{$\gamma_\bK$ describes the growth of the amplitude from the instability, and its appearance can be formally proven by keeping the imaginary part of the dielectric tensor $\varepsilon^{\mu\nu}$ as shown in Ref.~\cite{Fiorillo:2024uki}.}
This equation is understood along the ray generated by the eikonal evolution $\bK(t)$. The term $U\cdot K$ appears because we have absorbed the phase $e^{i S}$ in the definition of $\Psi_\bK$. \edit{For static seeding, a flavomon moving along a given ray, in a small time interval $\delta t$, acquires a relative phase equal to $-U\cdot K \delta t$, since both its clock and location are changing.}
Generally, the phase relation between the seeding term and the rapidly varying wave amplitude must be followed in this approach. The focusing term proportional to $\boldsymbol{\nabla}\cdot\bv_{\rm gr}$ describes the enhancement in the amplitude when several rays are coming together, and its eventual divergence at a caustic surface where multiple rays touch and geometrical optics fails as a wave description. 

To achieve a wave description, we need to know how different rays approach each other, encoded in the function $P_{ij}=\partial K_i/\partial r_j$; indeed
\begin{equation}
    \boldsymbol{\nabla}\cdot\bv_{\rm gr}=\frac{\partial^2\Omega}{\partial r_i \partial K_i}+\frac{\partial^2\Omega}{\partial K_i \partial K_j}\frac{\partial K_i}{\partial r_j}.
\end{equation}
For that, we need to use the equations of motion
\begin{equation}
    \frac{\partial K_i}{\partial t}+\frac{\partial \Omega}{\partial K_j}\frac{\partial K_i}{\partial r_j}=-\frac{\partial \Omega}{\partial r_i};
\end{equation}
upon differentiation this gives a closed equation for $P_{ij}$
\begin{eqnarray}
   \kern-1.5em \frac{\partial P_{ij}}{\partial t}+\frac{\partial^2 \Omega}{\partial K_k \partial r_j}P_{ik}+\frac{\partial^2\Omega}{\partial K_k \partial K_l}P_{ik} P_{lj}+\frac{\partial \Omega}{\partial K_k}\frac{\partial P_{ij}}{\partial r_k}
    \nonumber \\ 
    =-\frac{\partial^2\Omega}{\partial r_i \partial r_j}-\frac{\partial^2 \Omega}{\partial r_i \partial K_k}P_{kj}.
\end{eqnarray}
For an initially plane wave $P_{ij}(t=0)=0$, and we see from this equation that the wave develops curvature, and a non-vanishing $P_{ij}$, in proportion to the second derivative of $\Omega$. Therefore, in a constant linear gradient we may neglect this effect.

When the amplitude of the unstable mode has grown significantly, the seeding term becomes irrelevant. In this case, the phase of the amplitude need not be followed and we can revert to the flavomon occupation number: if we introduce then $N(\br, \bK; t)=|\Psi_\bK|^2 \delta(\bK - \boldsymbol{\nabla} S)$ summed over all the WKB solutions, by taking the derivative we find
\begin{equation}
    \left(\partial_t+\frac{\partial \Omega}{\partial \bK}\cdot \partial_\br-\frac{\partial \Omega}{\partial \br}\cdot \partial_\bK\right)N(\br,\bK;t)=2\gamma N(\br,\bK;t),
\end{equation}
describing the free transport of flavomons. The focusing term vanishes, as it should since transport conserves the volume in phase space by Liouville's theorem. We can see this explicitly; if we call $\mathcal{L}=d/dt$ the Liouville operator on the left-hand side, we have $\mathcal{L}|\Psi_\bK|^2(\br,t)=-\boldsymbol{\nabla}\cdot\bv_{\rm gr} |\Psi_\bK|^2(\br,t)$ and, calling $\boldsymbol{\nabla}S=\tilde{\bK}(\br,t)$
\begin{eqnarray}
    \kern-2em\mathcal{L}\,\delta[\bK-\tilde{\bK}(\br,t)]&=&\frac{\partial \delta(\bK-\tilde{\bK}(\br,t))}{\partial K_i}
    \nonumber\\
    \kern-2em&\times&\left[-\frac{\partial\Omega}{\partial r_i}-\frac{\partial \tilde{K}_i}{\partial t}-\frac{\partial\Omega}{\partial K_j}\frac{\partial \tilde{K}_i}{\partial r_j}\right]\!.
\end{eqnarray}
If we transport the derivative on the function in front of the delta, we finally find
\begin{eqnarray} 
    \kern-2em&&\mathcal{L}\,\delta[\bK-\tilde{\bK}(\br,t)]
    \nonumber\\
    \kern-2em&&=\,\delta[\bK-\tilde{\bK}(\br,t)]\left[\frac{\partial^2\Omega}{\partial r_i\partial K_i}+\frac{\partial^2\Omega}{\partial K_j\partial K_i}\frac{\partial \tilde{K}_i}{\partial r_j}\right]
    \nonumber\\ 
    \kern-2em&&=\,\boldsymbol{\nabla}\cdot \bv_{\rm gr}\,\delta[\bK-\tilde{\bK}(\br,t)],
\end{eqnarray}
so the two terms cancel exactly.

\textbf{\textit{Covariant equations of motion}}---In the main text, we have heuristically argued that gravitational effects on flavomon propagation are small so that their geometric-optics rays are mostly determined by the refractive weak-interaction effect of the matter background. To substantiate this finding, we here express the flavomon equations of motion in a general covariant form.

For simplicity, we stick to the case of neutrinos only without antineutrinos; it is straightforward to extend the result to antineutrinos. The covariant equations of motion then read

\begin{equation}
    p\cdot \partial \rho_\sp=-i[\Lambda u\cdot p + \frac{M^2}{2}+\sqrt{2}\GF p\cdot \rho,\rho_\sp].
\end{equation}
Here $p^\mu=(|\bp|,\bp)$ is the neutrino four-momentum, and the total four-current of the neutrinos is
\begin{equation}
    \rho^\mu=\int \frac{d^3\bp}{(2\pi)^3 E_\bp}\rho_\sp p^\mu;
\end{equation}
the off-diagonal component of this matrix represents the flavomon field.

The transport equation can be promoted to a generally covariant form by adding the drift terms of neutrino momenta
\begin{equation}
    p\cdot \partial \rho_\sp-\Gamma^i_{\alpha\beta} p^\alpha p^\beta \partial_{p_i}\rho_\sp=-i[\Lambda u\cdot p + \frac{M^2}{2}+\sqrt{2}G_F p\cdot \rho,\rho_\sp],
\end{equation}
where $\Gamma^\mu_{\nu\alpha}$ are the Christoffel symbols. 

From here, it is easy to deduce the covariant equation for $\psi_\sp$
\begin{eqnarray}
    p\cdot\partial \psi_\sp -\Gamma^i_{\alpha \beta}p^\alpha p^\beta \partial_{p_i}\psi_\sp&=&i(\lambda^\mu+D^\mu)p_\mu \psi_\sp -i\psi^\mu p_\mu D_\sp 
    \nonumber\\
        &&{}-i\frac{\delta m^2}{2} (c_V \psi_\sp+s_V D_\sp).
    \nonumber\\
\end{eqnarray}

We can now deduce directly the eikonal equation. First, we note that the transport term proportional to the gravitational field $\Gamma^i_{\alpha\beta}$ can be neglected. While gravitational drifts generally do impact the neutrino momentum distribution, their order of magnitude in the flavor transport equation is much smaller than all the other terms. Indeed, the first term $p\cdot \partial \psi_\sp\sim E_\nu \omega \psi$ where $E_\nu$ is the typical neutrino energy, $\omega$ is the typical flavomon frequency, and $\psi$ is the amplitude of the flavomon field. Instead, the second term is in order of magnitude $E_\nu \psi \phi/R$, where $\phi\sim GM/R\sim 0.1$ is the gravitational potential, and $R$ is the radius. Thus, this term is certainly negligible by many orders of magnitude in the flavomon dispersion relation, because it changes over macroscopic length scales. Even in the amplitude evolution equation, it can still be dropped, as it is suppressed by the relatively small potential $\phi$.

If we drop this term, we find that the dispersion relation of flavomons is determined exactly as in the absence of gravitational field, except for the presence of the metric. The latter can be removed by using the local values for the frequency and momentum. Let us specialize to a weak Newtonian field
\begin{equation}
    ds^2=\alpha^2 dt^2-\frac{|d\br|^2}{\alpha^2}
\end{equation}
with $\alpha=1+\phi$ and $\phi=-GM/R\ll1$. Then we can define the local frequency and wavevector
\begin{equation}
    \Omega_{\rm loc}=-\frac{1}{\alpha(\br)}\frac{\partial S}{\partial t}=\frac{\Omega_\infty}{\alpha(\br)},\;\bK_{\rm loc}=\alpha(\br)\frac{\partial S}{\partial \br}=\alpha(\br) \bK.
\end{equation}
For the coordinate frequency we have used the notation $\Omega_\infty$, which we explain in a moment. If the neutrino momenta $p^\mu$ are also expressed in terms of their local values, as they anyway are usually extracted from a SN simulation, then the eikonal equation at any given point gives simply the same dispersion relation as in the absence of the field $\Omega_{\rm loc}(\br,\bK_{\rm loc})$. The dependence on the radius comes from the effective inhomogeneity of the matter density, and to a smallest degree from the inhomogeneous neutrino distributions. However, as a flavomon traverses different radii, what is effectively conserved is its coordinate frequency; this is a direct consequence of Noether's theorem, since the eikonal equation does not contain time explicitly. Therefore, finally, the Hamiltonian for the flavomon is
\begin{equation}
    \mathcal{H}(\br,\bK)=\alpha(\br)\Omega_{\rm loc}[\br, \alpha(\br) \bK]=\Omega_\infty
\end{equation}
and remains conserved as it should for propagation in a static potential. The Hamilton equations then give directly the flavomon equations of motion
\begin{eqnarray}
    \frac{d\br}{dt}&=&\alpha^2\bv_{\rm gr},\\ \nonumber\frac{d\bK}{dt}&=&-\Omega_{\rm \infty}\frac{\partial\log\alpha}{\partial \br}-\alpha \left.\frac{\partial \Omega_{\rm loc}}{\partial \br}\right|_{\bK_{\rm loc}}-\alpha \bv_{\rm gr}\cdot \bK \frac{\partial \alpha}{\partial \br}.
\end{eqnarray}
We explicitly stress that the group velocity $\bv_{\rm gr}$ is evaluated at the local value of wavevector, and the derivative of $\Omega_{\rm loc}/\partial \br$ is taken at a fixed local wavevector.
For weak gravitational fields we may then take only the dominant gradient corrections and neglect all small terms, obtaining
\begin{equation}
    \frac{d\br}{dt}=\bv_{\rm gr},\quad\frac{d\bK}{dt}=-\Omega_\infty \frac{\partial \phi}{\partial \br}-\frac{\partial \Omega_{\rm loc}}{\partial \br}(1+\phi);
\end{equation}
the last term was neglected since $\bv_{\rm gr}\cdot \bK\ll \lambda\sim \Omega_{\rm \infty}$. The first term describes the gravitational drift; however, since $\phi \sim 0.1$, even this term gives negligible corrections in comparison with $\partial\Omega_{\rm loc}/\partial \br$. 

This estimate shows that gravitational drift of flavo\-mons is negligible in comparison with its weak drift. The situation is characteristically different from neutrinos, since for the latter, the weak interaction potential energy is orders of magnitude smaller than their total kinetic energy, so gravity exerts the vastly predominant force on their trajectories.

\onecolumngrid


\end{document}